\documentclass[twocolumn,prb,amsmath,amssymb,floatfix]{revtex4}

\usepackage{graphics}
\usepackage{graphicx}
\usepackage{dcolumn}
\usepackage{bm}
\usepackage{subfigure}
\usepackage{slashed}
\usepackage{bbold}
\usepackage{feynmf} 

\def\B{\langle }
\def\K{\rangle }

\def\eps{\epsilon}
\def\om{\omega}

\def\be{\begin{equation}}
\def\ee{\end{equation}}
\def\bea{\begin{eqnarray}}
\def\eea{\end{eqnarray}}
\def\bse{\begin{subequations}}
\def\ese{\end{subequations}}
\def\bc{\begin{center}}
\def\ec{\end{center}}

\catcode`,\active

\catcode`\,12

\begin{document}

\title{One-dimensional interacting electrons beyond the Dzyaloshinskii-Larkin theorem}

\author{S. Teber}
\affiliation{Laboratoire de Physique Th\'eorique et Hautes Energies, Universit\'e Pierre et Marie Curie, 4 place Jussieu, 75005, Paris, France}

\date{\today}

\begin{abstract}
We consider one-dimensional (1D) interacting electrons beyond the Dzyaloshinskii-Larkin theorem, {\it i.e.},
keeping forward scattering interactions among the electrons but adding a non-linear correction to the electron dispersion relation.
The latter generates multi-loop corrections to the polarization operator and electron self-energy 
thereby providing a variety of inelastic processes affecting equilibrium as well as non-equilibrium properties of the 1D system.  
We first review the computation of equilibrium properties, {\it e.g.}, the high frequency part of the dynamical structure factor
and corrections to the electron-electron scattering rate. On this basis, microscopic equilibration processes are identified and
a qualitative estimate of the relaxation rate of thermal carriers is given.
\end{abstract}

\maketitle


\section{Introduction} 

In systems of interacting electrons in one dimension certain physical phenomena such as 
damping of the elementary excitations and energy relaxation, are beyond the reach of the conventional Tomonaga-Luttinger (TL) 
model; see, {\it e.g.}, Ref.~[\onlinecite{ISG-2011}] for a review. 
This may be understood from the fact that, for such a model, the spectrum of the interacting electrons is in exact one-to-one 
correspondence with the one of free bosons.~\cite{Tomonaga-Luttinger-Mattis+Lieb} In diagrammatic terms this correspondence 
is generally referred to in the literature as the Dzyaloshinskii-Larkin (DL) theorem;~\cite{DL-1974,DiCM-1991+1993} see also 
Refs.~[\onlinecite{Review:Soloyom,Review:Voit,Review:Maslov,Book:Giamarchi}] for reviews. 
Inelastic processes responsible for a finite life-time of the elementary excitations and 
relaxation in one dimension may, however, be accessed by abandoning Tomonaga's approximation 
of linearizing the electron dispersion relation~\cite{Schick-1968} 
at the expense of losing the exact solvability of the model. 
A number of methods have been developed, especially during the last decade, to address these issues 
thereby generating a considerable amount of work; see Ref.~[\onlinecite{ISG-2011}] and references therein.
The purpose of this paper is to emphasize the importance of yet another approach based on the direct 
analysis of the lowest order multi-loop diagrams which are beyond the DL theorem. 
A review of the method for the computation of equilibrium properties
will first be presented. On this basis,
microscopic processes responsible for the equilibration of one-dimensional (1D) electrons will be identified.
The analysis will be qualitative but will nevertheless allow us to give an estimate of the relaxation rate of 
thermal carriers in quantum wires. This estimate agrees with the recent literature on the subject.~\cite{ML-2011,LMRM-2011,KGvO-2010}

\section{Dispersion non-linearity and interactions} 

\begin{figure}
  \includegraphics{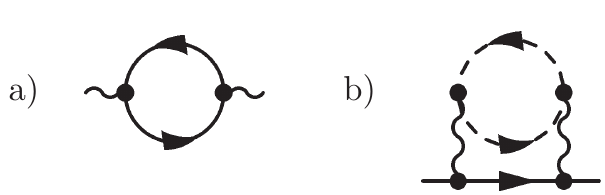}
  \caption{\label{fig:pi-one-loop+sigma-2-loop}
    (a) One-loop polarization diagram and (b) two-loop electron self-energy (for a right mover) for the scattering process $g_2$.}
\end{figure}

The TL model is exactly solvable thanks to two crucial 
approximations related to the shape of the electron dispersion relation and 
to the nature of the scattering processes between the electrons.
The former consists in linearizing the electron dispersion relation: 
$\xi_k^{\pm} = \pm v_F (k \mp k_F)$, where $\pm k_F$ are the Fermi points,
$\pm$ correspond to the chiralities of the fermions ($+$ for right movers 
near $+k_F$ and $-$ for left movers near $-k_F$), and $v_F$ is the Fermi velocity. 
As will be seen shortly, it is this approximation which is responsible for the fact that 
low-energy excitations are free bosons, {\it i.e.}, quantized sound 
waves~\cite{Bloch-1933+Ferrell-1964} of the 1D system. The properties of these collective 
modes are encoded in the density-density correlation function, 
$\Pi(x,\tau)= \B \rho(x,\tau) \rho(0,0) \K$, where $\rho$ is the density operator.
For non-interacting electrons, this correlation function reduces to a 1-loop diagram, see 
Fig.~\ref{fig:pi-one-loop+sigma-2-loop}(a). In imaginary time and in the linear spectrum approximation it reads~\cite{Review:Soloyom,Review:Voit,Review:Maslov,Book:Giamarchi}
\bea
\label{pi-FF}
\Pi_\pm^{(0)}(i\om,q) =  \pm \frac{1}{2\pi}\frac{q}{i\om \mp v_Fq},
\eea
where $\om$ is the energy of the boson of momentum $q$ and $\pm$ refer to the chirality of the fermions. 
The total polarization operator is given by: $\Pi^{(0)} = \Pi_+^{(0)} + \Pi_-^{(0)}$.
The imaginary part of the retarded polarization operator defines the dynamical structure factor (DSF):
$S(\om, q) = - \frac{2}{\pi}\,\Im \Pi^R(\om,q)$.
Combining this definition with Eq.~(\ref{pi-FF}) yields
\be
S^{(0)}(\om, q) = \frac{|q|}{\pi}\,\delta(\om - v_F|q|),
\label{DSF-FF}
\ee
where we assume that $\om>0$ for simplicity. 
Equation~(\ref{DSF-FF}) together with the $f$-sum rule~\cite{Book:Nozieres}
\be
\int_0^\infty {{d\om}} ~\om~ S(\om,q) = {\bar n}~\frac{q^2}{2m},
\label{FSR}
\ee
where ${\bar n}$ is the average electron density along the wire,
show that indeed the free boson of energy $\om = v_F|q|$ exhausts the $f$-sum rule. 

A coupling between the collective modes can be achieved by taking into account
small deviations from linear dispersion. Away from half-filling, the first correction is due
to the curvature of the electron dispersion relation: $\xi_k^{\pm} = \pm v_F (k \mp k_F) + (k\mp k_F)^2 /2m$,
which introduces an additional parameter, $m$, the band mass such that: $k_F \approx m v_F$, at small fillings. 
Including this non-linearity, the free fermion polarization operator and DSF (at $T=0$) become
\begin{subequations}
\label{FF+m}
\bea
&&\Pi^{(0)}(i\om,q)= \frac{m}{2\pi q} \ln{ \left[\frac{(i \om)^2 - (\om_-)^2}{(i\om)^2 - (\om_+)^2} \right]},
\label{pi-FF+m} \\
&& S^{(0)}(\om,q) = \frac{m}{2 \pi |q|} \left[ \Theta (\om - \om_-) - \Theta (\om - \om_+) \right],
\label{DSF-FF+m}
\eea
\end{subequations}
where the two-parametric family appears: $\om_\pm (q) = v_F|q| \pm q^2/2m$.
The spectral weight now spreads over a range of frequencies: $\om_-(q) < \om < \om_+(q)$. 
The latter corresponds to a continuum of single (particle-hole) pair excitations~\cite{Book:Nozieres}
which is responsible for the strong damping of the sound mode.
From Eq.~(\ref{DSF-FF+m}) we see that the $f$-sum rule of Eq.~(\ref{FSR}) is now exhausted by 
the single-pair excitations.

\begin{figure}
  \includegraphics{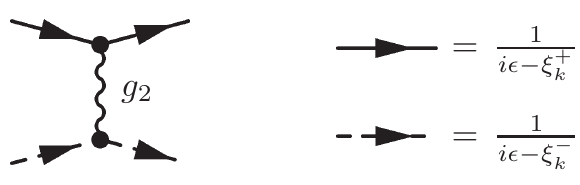}
  \caption{\label{fig:g2}
    Scattering process $g_2$ and bare chiral propagators
    (solid line for right movers, dashed line for left movers, wiggly line for the interaction)
    with non-linear dispersion: $\xi_k^{\pm} = \pm v_F (k \mp k_F) + (k\mp k_F)^2 /2m$.}
\end{figure}

The second approximation made in the TL model is that only forward-scattering
processes associated with small momentum transfers are taken into account.
The corresponding dimensionless coupling constants are noted $g_4$ and $g_2$ for processes
involving the scattering of electrons of the same and of different chiralities, respectively,
see Fig.~\ref{fig:g2} for $g_2$. On the other hand, backscattering ($g_1$) and umklapp scattering ($g_3$),
which are characterized by large momentum transfers are neglected.
Under these assumptions, $g_4$ and $g_2$ are invariant under the renormalization group flow
and the TL model appears as an infrared fixed point of a more general
class of ``$g$-ology" models.~\cite{Review:Soloyom,Review:Voit,Review:Maslov,Book:Giamarchi} It describes a 1D metal; {\it i.e.}, excitations are gapless.
The difference with the 3D (Fermi liquid) case may be seen~\cite{Review:Soloyom,Review:Voit,Review:Maslov,Book:Giamarchi} perturbatively from the electron-electron scattering rate
and quasi-particle residue: $\tau_k^{-1} = - \Im \Sigma(\xi_k,k)$ and $z_k^{-1} = [1-\partial \Re \Sigma(\eps,k)/\partial \eps]_{\eps=\xi_k}$, 
respectively, where $\Sigma$ is the electron self-energy; see Fig.~\ref{fig:pi-one-loop+sigma-2-loop}(b)
for the $g_2$ process. The latter determine the dressed single particle Green's function, $G(k,\eps) = 1/(\eps - \xi_k -\Sigma(k,\eps))$ and
the spectral function, $A(k,\eps) = -\frac{1}{\pi}\,\Im G(k,\eps)$.
For a Fermi liquid (FL), $\Im \Sigma \sim -g^2\,{\mathrm{max}}\{\eps^2,T^2\}/\eps_F$ 
and $\Re \Sigma(\eps,k) \sim g^2 \eps$, implying the existence of well-defined quasi-particles:
$\tau_{FL}^{-1} \sim g^2\,{\mathrm{max}}\{\eps^2,T^2\}/\eps_F$, $z_{FL} \sim 1$ with a Lorentzian spectral line-shape.
In 1D on the other hand, $\Im \Sigma \sim -g_2^2\,{\mathrm{max}}\{|\eps|,T\}$ and 
$\Re \Sigma \sim g_2^2 \,\eps \,\ln ({\mathrm{max}}\{|\eps|,T\}/\eps_F)$.
The scattering rate is therefore of the order of the energy of the particle:
$\tau_{TL}^{-1} \sim g_2^2\,{\mathrm{max}}\{|\xi_k|,T\}$, and 
the quasi-particle residue vanishes logarithmically at low energies:
$z_{TL}^{-1} \sim 1-g_2^2 \ln ({\mathrm{max}}\{|\xi_k|,T\}/\eps_F)$, as in a marginal Fermi liquid.~\cite{Varma-1989}
At very weak couplings, $g_2 \ll 1$, fermionic quasi-particles survive above an exponentially small energy
scale:  $T > \eps_F \exp(-1/g_2^2)$. However, the linear dependence on energy of the scattering rate
implies that the spectral line shape is non-Lorentzian with high-frequency tails scaling as $A(k,\eps) \sim 1/{\mathrm{max}}\{|\eps|,T\}$,
in contrast to the Fermi-liquid case.

Concerning the bosonic correlation function, the combined assumptions of forward-scattering interactions and linear dispersion
imply that there is no higher order correction to the lowest order diagram of Fig.~\ref{fig:pi-one-loop+sigma-2-loop}(a).
In this sense the TL model is exact at one loop. This is the essence of the DL (or
loop-cancellation) theorem which allows for an exact computation of all correlation functions at arbitrary 
interaction strength.~\cite{DL-1974,DiCM-1991+1993} Beyond weak coupling, single-particle excitations disappear and the system
becomes a non-Fermi liquid. Concerning the collective mode, interactions renormalize
its velocity, $v \not= v_F$, and bring a prefactor, $\gamma \not= 1$,
fixed by the $f$-sum rule in the expression of the DSF, which may then be written as
$S_{TL}(\om, q) = \gamma |q|\,\delta(\om - v|q|)$. Bosons therefore remain free
and the boson-fermion mapping implies that there is no relaxation in the TL model.

As anticipated in the Introduction, relaxation requires both dispersion non-linearity and interactions among the electrons.
The former is irrelevant in the renormalization group sense and is not expected to affect
the infrared properties of the model. But when combined with interactions among the fermions
the DL theorem breaks down and multi-loop corrections appear providing a variety of collision
processes. Presently, no consistent resummation scheme exists to take into account these processes 
in a non-perturbative way. In the following we will therefore simply focus on the lowest order non-trivial 
corrections beyond the DL theorem and relate such corrections to the equilibration of 1D interacting fermions.

\section{Perturbation theory}

We assume that both the interactions among the electrons and the
deviations from linear dispersion are weak and take them into account in perturbation theory. 
For simplicity, we focus on spinless fermions, {\it i.e.}, fully spin-polarized electrons. 
In this case there is no distinction between $g_1$ and $g_2$ processes. 
The combinations $g_4$ and $g_2-g_1$ are then invariant under the renormalization group flow and 
have to be taken into account. If we assume that electrons interact via the realistic 3D Coulomb interaction, {\it i.e.},
$V(q) = -e^2 \ln(qa)$, where $q$ is the 1D momentum transfer and $a$ the width of the wire,
these coupling constants depend on momentum only logarithmically. In the following we will neglect this
marginal dependence keeping in mind, however, that $g_2-g_1 = [V(0) - V(2k_F)]/2\pi v_F \rightarrow g_2 \not=0$,
in the logarithmic approximation. Then, to logarithmic accuracy, direct and exchange processes involving 
$g_4$ cancel each other and we are left with the momentum-independent process $g_2$, see Fig.~\ref{fig:g2}. 
This considerably reduces the number of diagrams that have to be considered, thereby simplifying the diagrammatic study while keeping track of
the most important microscopic processes.

\begin{figure}
  \includegraphics{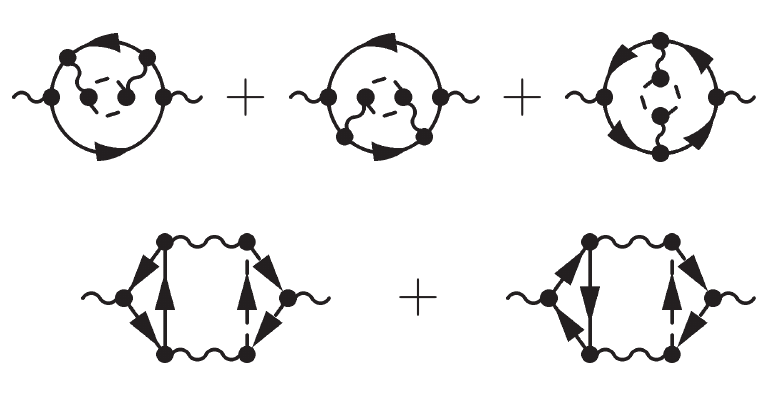}
    \caption{\label{fig:pi-3loop}
    Three-loop corrections to the polarization diagram. Top: $\Pi_{++}^{(2)}$ (and equivalently for $\Pi_{--}^{(2)}$ upon exchanging
solid and dashed lines). Bottom: $\Pi_{\pm \mp}^{(2)}$.}
\end{figure}

With these assumptions in hand, the lowest order non-trivial diagrams that need to be considered are the three-loop diagrams displayed in Fig.~\ref{fig:pi-3loop}.
For clarity they have been separated into three groups with obvious notations: ``chiral'' $\Pi_{\pm \pm}$ diagrams as well
as a group of ``mixed'' $\Pi_{\pm \mp}$ diagrams generally referred to as the Aslamasov-Larkin diagrams in the literature.
The sum of all these contributions yields the lowest non-trivial correction, due to both dispersion non-linearity and
interactions, to the polarization operator:
$\Pi^{(2)} = \Pi_{++}^{(2)} + \Pi_{--}^{(2)} + 2\Pi_{+-}^{(2)}$,
and the corresponding DSF. These diagrams have been computed to the lowest order in $1/m$ in
Ref.~[\onlinecite{Teber-2006}]. For clarity, we reproduce here the expression of the imaginary part of the corresponding retarded
polarization operators:
\begin{subequations}
\label{pi-3loop-RLM}
\bea
&&\Im \Pi_{\pm \pm}^{(2)R}(\om,q) = - \frac{g_2^2 \pi}{4v_F}\,\frac{q^2}{(m v_F)^2}\,\frac{\om \pm v_Fq}{\om \mp v_Fq}\,{\mathcal{F}(T;\om,q)},
\label{R and L} \\
&&\Im \Pi_{\pm \mp}^{(2)R}(\om,q) = + \frac{g_2^2 \pi}{4v_F}\,\frac{q^2}{(m v_F)^2}\,{\mathcal{F}(T;\om,q)},
\label{M}
\eea
\end{subequations}
where, for simplicity, a numerical coefficient has been absorbed in the coupling constant $g_2$ and the
thermal factor is given by:
${\mathcal{F}} = 1/(1 - e^{-\frac{\om - v_Fq}{2T}}) - 1/(1 - e^{\frac{\om + v_Fq}{2T}})$.
From Eqs.~(\ref{pi-3loop-RLM}) the three-loop correction to the DSF reads
\be
S^{(2)}(\om,q) = \frac{g_2^2}{v_F}\,\left(\frac{q^2}{m}\right)^2\,\frac{1}{\om^2 - (v_Fq)^2}\,{\mathcal{F}(T;\om,q)}.
\label{DSF-3-loop}
\ee
At zero temperature, ${\mathcal{F}(T;\om,q)} = \Theta(\om - v_Fq)$ for $\om>0$.
Equation~(\ref{DSF-3-loop}) therefore corresponds to a continuum of two-pair excitations which extends beyond
the single-pair continuum, at $\om \gg v_F |q|$. Thanks to the high-frequency cancellation brought by the
diagram of Eq.~(\ref{M}) this two-pair continuum has a vanishing contribution to the spectral weight and the 
sum rule, Eq.~(\ref{FSR}), is still exhausted by the single-pair excitations. This perturbative
result has limitations as $\om \rightarrow v_F|q|$ where non-perturbative approaches are 
required~\cite{Rozkhov,Pustilnik,Pereira} but nevertheless agrees with similar results obtained by different 
methods~\cite{Pustilnik,Pereira,Aristov,Kopietz}.

\begin{figure}
  \includegraphics{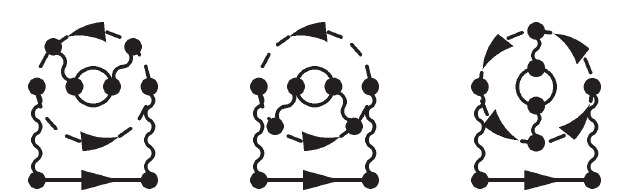}
    \caption{\label{fig:sigma-4loop}
    Four-loop corrections to the self-energy of a right-moving fermion 
    (and equivalently for a left mover upon exchanging solid and dashed lines). }
\end{figure}

On the basis of these three-loop diagrams we can estimate corrections to the electron-electron scattering rate beyond the DL theorem:
$\tau^{-1} = \tau^{(2)-1} + \tau^{(4)-1} +...$, where $\tau^{(2)-1} \equiv \tau_{TL}^{-1} \sim g_2^2\,T$ for a thermal excitation 
as already discussed. The next correction is given by the imaginary part of the 4-loop fermion self-energy,
Fig.~\ref{fig:sigma-4loop}. Due to constraints imposed by the $g_2$ process the Aslamasov-Larkin diagrams do not contribute to the 
fermion self-energy. The remaining diagrams describe the scattering of two (particle-hole) pairs of different chiralities
and therefore do carry information about the life-time of the excitations. The latter is estimated to be
\bea
&&\tau^{(4)-1} \sim g_2^4\,\frac{T^3}{\eps_F^2},
\label{rate}
\eea
up to a numerical factor. This correction brings back the Fermi energy scale but
does not follow the usual Fermi-liquid-like scaling: $T^2 / \eps_F$. 
From dimensional arguments we understand the cubic energy-dependence as the only possible scaling to compensate for the 
square of the Fermi energy appearing from the polarization part. 

\section{Kinetics and thermalization rate} 

Following Refs.~[\onlinecite{ML-2011,KGvO-2010}] we now consider the hypothetical situation where 
the system is prepared in an out-of-equilibrium state characterized by a small temperature difference in the 
distribution of right and left movers. The scattering process $g_2$ brings back the system to equilibrium.
In the following, we give evidence that the corresponding thermalization rate is given by Eq.~(\ref{rate}).

\begin{figure}
  \includegraphics{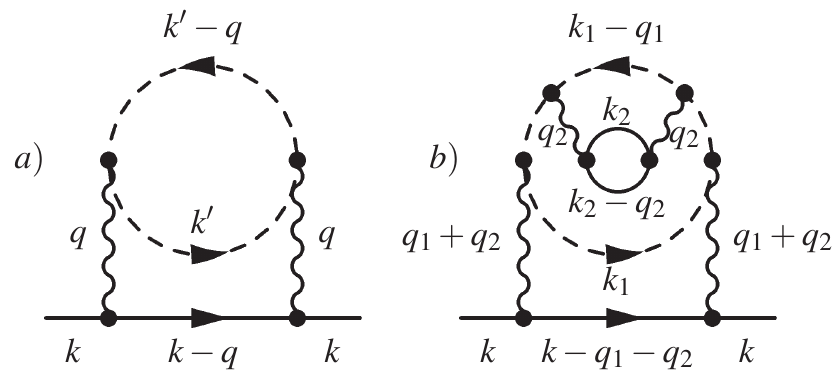}
    \caption{\label{fig:sigma-kinetic}
    Comparison between (a) two-loop and (b) one of the four-loop diagrams of Fig.~\ref{fig:sigma-4loop}. Diagram (a) yields a four-fermion collision integral.
    Diagram (b), together with the other diagrams of Fig.~\ref{fig:sigma-4loop}, yields a six-fermion collision integral along the cut passing through five particle lines.}
\end{figure}

A rigorous approach to kinetics should in principle be based on the non-equilibrium Green's function technique;
see, {\it e.g.}, Refs.~[\onlinecite{Review:NEGF}] for reviews. Here, we shall follow the basic steps of the method on a qualitative level.
This technique allows one to construct a quantum kinetic equation (QKE) for a generalized distribution function: $n(\eps, k, t)$.
For a Fermi liquid the QKE reduces to a Boltzmann equation: $\partial_t n_k = - I(n_k)$,
where $n_k \equiv n(\eps=\xi_k,k,t)$ is the electron distribution function and all the relaxation mechanisms are in the collision integral, $I(n)$.
For 1D interacting electrons the situation is highly non-trivial. The first difficulty is that, as the interaction strength increases,
quasi-particles become ill-defined in 1D. To circumvent this difficulty, we restrict ourselves to the case of very weak
interactions and not too low temperatures where, as previously discussed, quasi-particles still exist.
However, the spectral function is non-Lorentzian: broader on-shell with part of the spectral weight transferred to
high frequency tails. As a consequence one cannot, in principle, neglect off-shell contributions:
$\eps \not= \xi_k$. For thermal carriers, we argue that the long time dynamics is dominated by the
on-shell contribution.~\footnote{From $\delta \eps \, \delta t \gtrsim 1$ and for an energy dispersion $\delta \eps \sim T$,
long time dynamics correspond to: $t \gg 1/T$. The off-shell contribution is dominated by 2-body collisions which affect the
dynamics on time scales: $\tau^{(2)} \sim 1 / g_2^2 T$. 
The on-shell contribution corresponds to 3-body collisions which affect the dynamics on time scales: $\tau^{(4)} \sim \eps_F^2 / g_2^4 T^3$.
The condition $\tau^{(4)} \gg \tau^{(2)}$ is clearly satisfied in a broad range of temperatures: $\eps_F \exp(-1/g_2^2) \ll T \ll \eps_F$, where $g_2 \ll 1$.
}
Focusing on these long time scales, which may be the easiest to access experimentally, we therefore neglect the off-shell contributions.
It is then well known~\cite{Review:NEGF} that, to lowest order in interactions, the collision integral has a four-fermion structure determined by
the two-loop self-energy diagram of Fig.~\ref{fig:sigma-kinetic}(a): $n_k n_{k'-q}(1-n_{k'})(1-n_{k-q})-n_{k'} n_{k-q}(1-n_{k})(1-n_{k'-q})$,
where the two terms correspond to ``in'' and ``out'' contributions. Energy and momentum conservation then impose: $\xi_k^+ + \xi_{k'-q}^- = \xi_{k'}^- + \xi_{k-q}^+$. In 1D
this yields: $q=0$ and hence the identical cancellation of ``in'' and ``out'' terms. We thus recover the fact that (on-shell) two-body
collisions do not contribute to relaxation in 1D.~\cite{Sirenko-1994}
The next correction, which is non-zero thanks to dispersion non-linearity, is given by the four-loop
diagrams of Fig.~\ref{fig:sigma-4loop}; see Fig.~\ref{fig:sigma-kinetic}(b) for a more detailed example. 
Tedious but straightforward calculations then show that
the corresponding collision integral has a six-fermion structure; {\it e.g.}, for the diagram of Fig.~\ref{fig:sigma-kinetic}(b): $n_k n_{k_1-q_1}n_{k_2-q_2}(1-n_{k_1})(1-n_{k_2})(1-n_{k-q_1-q_2})
-n_{k_1} n_{k_2}n_{k-q_1-q_2}(1-n_k)(1-n_{k_1-q_1})(1-n_{k_2-q_2})$. The diagrams of Fig.~\ref{fig:sigma-4loop} yield a non-vanishing contribution to the collision integral
even when the energy-momentum constraint is taken into account. On shell, such three-body collisions therefore provide the leading contribution
to the thermalization rate which scales as: $g_2^4\,T^3/\eps_F^2$, see Eq.~(\ref{rate}).

\section{Conclusion} 

A brief account has been given of the analysis of the lowest order non-trivial corrections beyond the DL theorem arising from interactions and dispersion non-linearity.
Focusing on the case of weak interactions and spinless fermions, a rigorous derivation of the three-loop corrections to the DSF has been reviewed.~\cite{Teber-2006}
The corresponding four-loop fermion self-energy diagrams were related to equilibration processes of 1D interacting fermions.
Qualitative arguments allowed us to estimate the temperature dependence of the thermalization rate, see Eq.~(\ref{rate}), in agreement 
with the literature on the subject;~\cite{ML-2011,LMRM-2011,KGvO-2010} see also Ref.~[\onlinecite{DGP-2012}] for recent progress.
The diagrammatic approach can be applied to the quantitative computation of other equilibrium properties of interest.
In light of recent literature~\cite{ML-2011,LMRM-2011,KGvO-2010,DGP-2012} the whole approach can be extended to, {\it e.g.},  various interaction potentials, to the $g_4$ 
process,~\footnote{The $g_4$ process is subject to an infrared catastrophe, see, {\it e.g.}, Ref.~\onlinecite{Review:Maslov} for a review, which is cured by curvature corrections.}
and, eventually, to the case of spinful fermions.



\begin{thebibliography}{99}

\bibitem{ISG-2011}
A.\ Imambekov et al., \rmp {\bf 84}, 1253 (2012).

\bibitem{Tomonaga-Luttinger-Mattis+Lieb}
S.-I.~Tomonaga, Prog.~Theor.~Phys. {\bf 5}, 544 (1950); J.\ M.\ Luttinger, J.\ Math.\ Phys.\ {\bf 4} 1154 (1963);
D.~C.~Mattis and E.~H.~Lieb, J.~Math.~Phys.\ {\bf 6}, 304 (1965).

\bibitem{DL-1974}
I.~E.~Dzyaloshinskii and A.~I.~Larkin, Sov.~Phys.-JETP {\bf 38}, 202 (1974).

\bibitem{DiCM-1991+1993}
C. Di Castro and W. Metzner, \prl {\bf 67}, 3852 (1991); \prb {\bf 47}, 16107 (1993).

\bibitem{Review:Soloyom}
J.~Sol\'oyom, Adv.\ Phys.\ {\bf 28}, 201 (1979). 

\bibitem{Review:Voit}
J.\ Voit, Rep.\ Prog.\ Phys.\ {\bf 58}, 977 (1995).

\bibitem{Review:Maslov}
D.\ Maslov, arXiv:cond-mat/0506035.

\bibitem{Book:Giamarchi}
T.~Giamarchi, {\sl Quantum Physics in One Dimension}, Oxford University Press, USA (2004).

\bibitem{Schick-1968}
M.\ Schick, Phys.~Rev.\ {\bf 166}, 404 (1968). 

\bibitem{ML-2011}
T.\ Micklitz and A.\ Levchenko, \prl {\bf 106}, 196402 (2011).

\bibitem{LMRM-2011}
A.\ Levchenko et al., \prb {\bf 84}, 115447 (2011).

\bibitem{KGvO-2010}
T.\ Karzig et al., \prl {\bf 105}, 226407 (2010).

\bibitem{Bloch-1933+Ferrell-1964}
F.~Bloch, ZS.\ Phys.\ {\bf 81} 363 (1933); R.\ A.\ Ferrell, \prl {\bf 13}, 330 (1964).

\bibitem{Book:Nozieres}
D.~Pines and P. Nozi\`eres, {\sl The Theory of Quantum Liquids}, Vol. 1, Perseus Books Publishing (1989).

\bibitem{Varma-1989}
C.~M.~Varma et al., \prl {\bf 63}, 1996 (1989).

\bibitem{Teber-2006}
S.~Teber, Eur.\ Phys.\ J.\ B {\bf 52}, 233 (2006).

\bibitem{Rozkhov}
A.~V.~Rozhkov, Eur.\ Phys.\ J.\ B {\bf 47}, 193 (2005);
\prb {\bf 74}, 245123 (2006);
\prb {\bf 77}, 125109 (2008).

\bibitem{Pustilnik}
M.\ Pustilnik et al., \prl {\bf 96}, 196405 (2006).

\bibitem{Pereira}
R.\ G.\ Pereira et al., \prl {\bf 96}, 257202 (2006). 

\bibitem{Aristov}
D.\ N.\ Aristov, \prb {\bf 76}, 085327 (2007).

\bibitem{Kopietz}
P.\ Pirooznia et al., Eur.\ Phys.\ J.\ B {\bf 58}, 291 (2007);
\prb {\bf 78}, 075111 (2008).

\bibitem{Review:NEGF}
J.~Rammer and H.~Smith, \rmp {\bf 58}, 323 (1986); 
A.~Kamenev, arXiv:cond-mat/0412296.

\bibitem{Sirenko-1994}
Y.~M.~Sirenko et al., \prb {\bf 50}, 4631 (1994).

\bibitem{DGP-2012}
A.~P.~Dmitriev, I.~V.~Gornyi and D.~G.~Polyakov, arXiv:1207.1089.

\end{thebibliography}
\end{document}